\documentclass[a4paper,11pt]{article}
\usepackage{epsf,latexsym,amsmath}
\usepackage{times}
\usepackage{amssymb}
\usepackage{graphicx}
\usepackage[all,poly]{xy}
\usepackage{rotating}
\input{epsf}
\xyoption{rotate}
\begin{document}


\begin{center}
{ \Large \bf  Quantum Tetrahedra} 
\end{center}
\vspace{24pt}

\begin{center}
{\sl Mauro Carfora}\\
Dipartimento di Fisica Nucleare e Teorica,
Universit\`a degli Studi di Pavia\\
 and INFN, Sezione di Pavia, 
via A. Bassi 6, 27100 Pavia (Italy);\\ 
E-mail: mauro.carfora@pv.infn.it 
\end{center}

\vspace{12pt}

\begin{center}
{\sl Annalisa Marzuoli}\\
Dipartimento di Fisica Nucleare e Teorica,
Universit\`a degli Studi di Pavia\\
 and INFN, Sezione di Pavia, 
via A. Bassi 6, 27100 Pavia (Italy);\\ 
E-mail: annalisa.marzuoli@pv.infn.it 
\end{center}

\vspace{12pt}

\begin{center}
{\sl Mario Rasetti}\\
Dipartimento di Fisica,
Politecnico di Torino\\
corso Duca degli Abruzzi 24, 10129 Torino (Italy)\\
and Institute for Scientific Interchange  Foundation,\\ 
viale Settimio Severo 75, 10131 Torino (Italy) \\
E-mail: mario.rasetti@polito.it 
\end{center}

\vspace{12pt}

\noindent {\bf Abstract}\\
 We discuss in details
 the role of Wigner $6j$ symbol as the basic
 building block unifying such different fields as  state sum models for quantum geometry,
topological quantum field theory, statistical lattice models and quantum computing.
The apparent twofold nature of the $6j$ symbol displayed in quantum field theory 
and quantum computing --a quantum tetrahedron and a
computational gate-- is shown to  merge together in a
unified quantum--computational $SU(2)$--state sum framework.

\vspace{12pt}

\noindent {\bf Keywords}: quantum theory of angular momentum; Wigner $6j$ symbol;
discretized quantum gravity; spin network quantum simulator

\vfill\eject

 \section{Introduction}
 \begin{figure*}[h]
\begin{center}
\includegraphics[bb= 0 0 540 470,scale=.4]{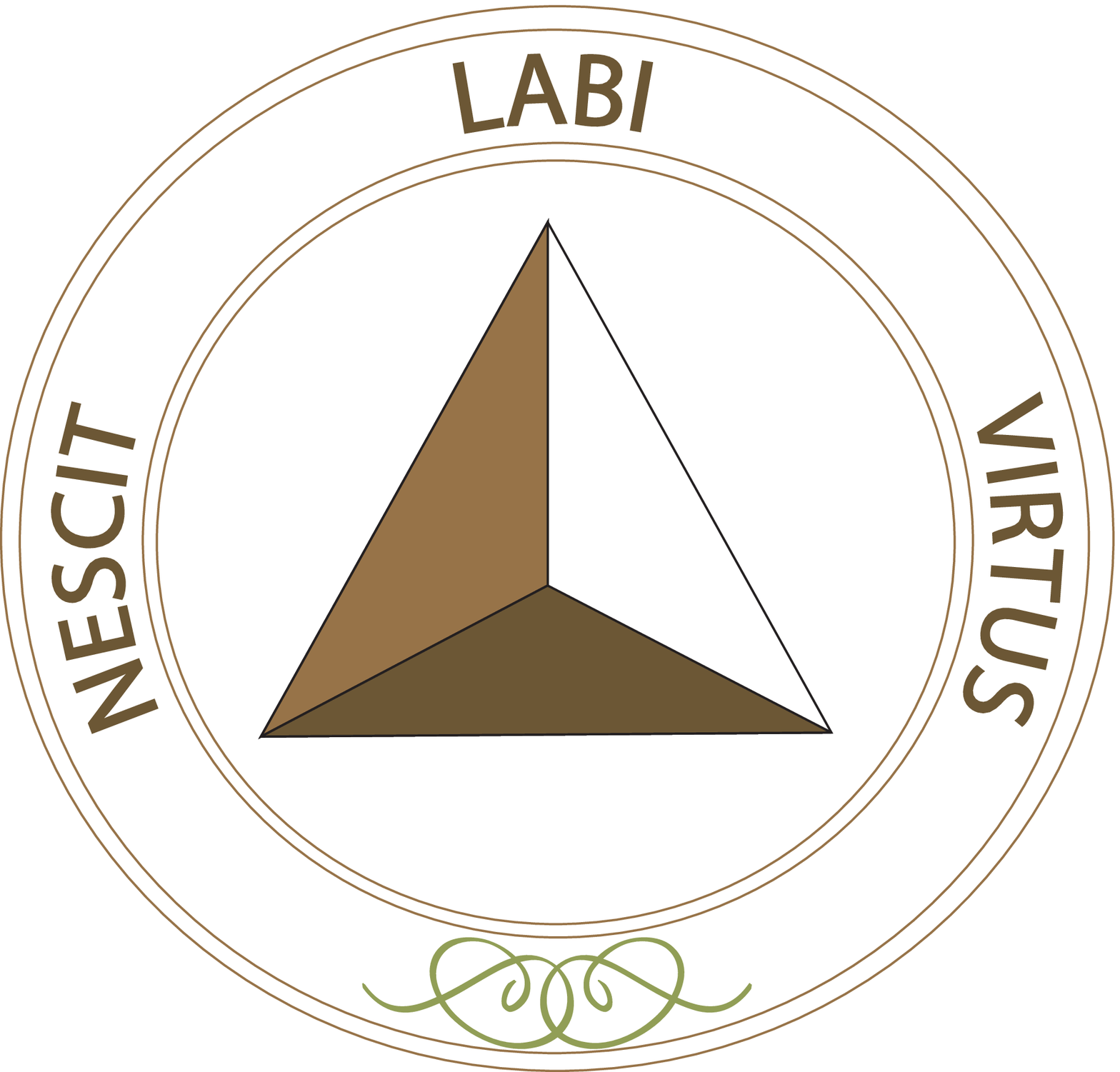}
\end{center}
\end{figure*}

\noindent The above illustration shows a variant woodcut printer's device 
on verso last leaf of a rare XVI century edition of Plato's Timaeus, 
(\emph{Divini Platonis Operum a Marsilio Ficino tralatorum, Tomus Quartus. 
Lugduni, apud Joan Tornaesium M.D.XXXXX}). The printer's device to the colophon 
shows a medaillon with a tetrahedron in centre, and the motto round the border: 
\emph{Nescit Labi Virtus}, Virtue cannot fail\footnote{a more pedantic rendering is: Virtue ignores 
the possibility of sliding down.}. This woodcut beautifully 
illustrates the role of the perfect shape of the tetrahedron in classical culture. 
The tetrahedron conveys such an impression of strong stability as to be considered 
as an epithome of virtue, unfailingly capturing us with the depth and elegance of its shape. 
However, as comfortable as it may seem, this time--honored geometrical shape smuggles energy 
into some of the more conservative aspects of Mathematics, Physics and Chemistry, 
since it is perceptive of where the truth hides away from us: the quantum world. 
As Enzo says, the geometry of the tetrahedron actually takes us on a trip pointing 
to unexpected connections between the classical and the quantum. He has indeed often  
entertained us with descriptions of open terrains of Physics and Chemistry which are bumpy, 
filled with chemical bonds and polyhedra, and which bend abruptly in unexpected directions. 
We do feel that, like any good adventure, it is not the destination, but what we unexpectedly 
found around the bend that counts. Thus, the story we wish to tell here is the story of what, 
together with Enzo, we found around the bend: the unfailing virtues of the quantum tetrahedron.

\vskip 2 cm

Our story starts by recalling that the (re)coupling theory of many 
$SU(2)$ angular momenta --framed mathematically in
the structure of the Racah--Wigner 
tensor algebra-- is the most exhaustive formalism in dealing  with interacting
many-angular momenta quantum systems \cite{BiLo8,BiLo9}. 
As such it has been over the years a common tool in
advanced applications in atomic and molecular physics, nuclear physics as well
as in  mathematical physics. Suffices here to mention in physical chemistry
the basic work of Wigner, Racah, Fano and others (see the collection of
reprints \cite{rep} and the Racah memorial volume quoted in \cite{PoRe} below)
as well as the recent book \cite{Ave} on topics covered in this special issue.

In the last three decades 
there has been also a deep interest in applying 
(extensions of) such notions and
techniques in the branch of theoretical physics
known as Topological Quantum Field Theory, as well
as in related discretized models for $3$--dimensional quantum gravity.
More recently the same techniques have been employed 
for establishing a new framework for quantum computing,
the so--called "spin network" quantum simulator.

In previous work in collaboration with Enzo
 \cite{AqBiFe}
 we have stressed the combinatorial properties of  Wigner $6j$ symbols
 (and of its generalizations, the $3nj$ symbols, see \cite{AnAqMa}) 
which stand at the basis of so many different fields
of research.

 The aim of the present paper is to discuss in details
 the apparent twofold nature of the $6j$ symbol displayed in quantum field theory 
and quantum computing,
and to convey the idea that these two pictures actually merge together.
In section 2 the $6j$ is looked at as a real "tetrahedron",
the basic {\em magic brick} in constructing $3$--dimensional
quantum geometries of the Regge type, while in section 3 it
plays the role of a  {\em magic box}, namely the elementary universal
computational gate in a quantum circuit model.
Thus the underlying physical models embody, at least  
in principle, the hardware of
quantum computing machines, while a quantum computer of this sort,
looked at as a universal, multi--purpose machine,
might be able to simulate "efficiently" any other
discrete quantum system. More remarks this topic
are postponed to the end of section 3, while
most  mathematical definitions and results
on Wigner $6j$ symbols needed in the previous
sections  are collected in Appendix A.

\section{Tetrahedra and 6j symbols in quantum gravity}

From a historical viewpoint the Ponzano--Regge
asymptotic formula for the $6j$ symbol \cite{PoRe}, 
reproduced in
(\ref{PRasymt}) of Appendix A.1, together with
the seminal paper \cite{Reg1} in which  "Regge Calculus"
was founded, are  no doubt at the basis of all
"discretized" approaches to General Relativity, 
both at the classical and at the quantum level.

In Regge's approach the edge lengths of a "triangulated" spacetime
are taken as discrete counterparts of the metric, a tensorial
quantity which encodes
the dynamical degrees of freedom of the gravitational field
and appears in the classical Einstein--Hilbert action for General Relativity
through its second derivatives combined in the Riemann scalar  curvature.
Technically speaking, a Regge spacetime
is a piecewise linear (PL) "manifold" of dimension $D$ dissected into
{\em simplices}, namely triangles in $D=2$, tetrahedra in $D=3$, 
4-simplices in $D=4$ and so on. Inside each simplex either an Euclidean or a Minkowskian metric
can be assigned: accordingly, PL manifolds obtained by gluing together
$D$--dimensional simplices acquire an overall $PL$ metric of Riemannian or Lorentzian
signature\footnote{Einstein's General Relativity corresponds to the
physically significant case of a $4$--dimensional spacetime endowed with
a smooth Lorentzian metric. However, models formulated in
"non--physical" dimensions such as $D=2,3$ turn out to be highly non trivial and very useful
in a variety of applications, ranging from conformal field theories
and associated statistical models in $D=2$ to the study of geometric
topology of $3$--manifolds. Moreover, the most commonly used quantization procedure of
such theories has a chance of being well--defined only when the underlying
geometry is (locally) Euclidean, see further remarks below.}.\\ 
Consider a particular triangulation $\mathcal{T}^D\,(\ell )\rightarrow \mathcal{M}^D$,
where $\mathcal{M}^D$ is a closed, locally Euclidean 
manifold of fixed topology and $\ell$ denotes collectively
the (finite) set of edge lengths of the simplices in  $\mathcal{T}^D$. 
The Regge action is given explicitly by (units are chosen such that the Newton constant $G$
is equal to $1$) 
\begin{equation}\label{ReAction}
S(\mathcal{T}^D\,(\ell )) \equiv S^D (\ell ) 
\,=\, \sum_{\sigma_i}\, \text{Vol}^{(D-2)}(\sigma_i) \, \epsilon_i \, ,
\end{equation}
where the sum is over $(D-2)$--dimensional simplices $\sigma_i \in$ $\mathcal{T}^D$
(called hinges or "bones"),
$\text{Vol}^{(D-2)}(\sigma_i)$ are their $(D-2)$--dimensional volumes expressed in terms of the edge lengths
and $\epsilon_i$ represent the deficit angles at $\sigma_i$. The latter are  defined, for each $i$,
as $2\pi - \sum_k \theta_{i,k}$, where $\theta_{i,k}$ are the dihedral angles between
pairs of $(D-1)$--simplices  meeting at $\sigma_i$ and labeled by some $k$. Thus a positive [negative or null]
value of the deficit angle  $\epsilon_i$ corresponds to a positive [negative or null] curvature to be assigned
to the bone $i$,
detected for instance by moving a $D$--vector along a closed path around the bone $i$ and measuring
the angle of rotation. Even  such a sketchy description of  Regge geometry should make it clear
that a discretized spacetime is flat (zero curvature) inside each $D$--simplex, while curvature is
concentrated at the bones which represent "singular"
subspaces. It can be proven that  the limit of the Regge action
(\ref{ReAction}) when the edge lengths become smaller and smaller gives the usual
Einstein--Hilbert action for a spacetime which 
is "smooth" everywhere, the curvature being distributed "continuously". Regge equations
--the discretized analog of Einstein field equations-- can be derived from the classical
action by varying it with respect to the dynamical variables, {\em i.e.} the set $\{\ell\}$ of
edge lengths of $\mathcal{T}^D\,(\ell )$, according to Hamilton principle of classical field theory 
(we refer to  \cite{WiTu} for a bibliography and brief review on Regge Calculus  
from its beginning up to the 1990's).\\

Regge Calculus  gave rise in the early  1980's 
to a novel approach to quantization of General Relativity
known as Simplicial Quantum Gravity (see  \cite{WiTu,AmCaMa,ReWi} and references therein). 
The quantization procedure
most commonly adopted is the Euclidean path--sum approach, namely a discretized
version of Feynman's path--integral describing
$D$--dimensional Regge geometries undergoing "quantum fluctuations"
(in Wheeler's words  a "sum over histories" \cite{MiThZu},
formalized for gravity in the so--called Hawking--Hartle prescription \cite{HaHa}).
Without entering into technical details, the discretized path--sum approach
turns  out to be very useful in addressing a number of conceptual open questions  in
the approach relying on the geometry of smooth spacetimes, although
the most significant improvements have been achieved for the $D=3$ case, which 
we are going to address in some details in the rest of this section.\\

Coming to the interpretation of  Ponzano--Regge asymptotic formula 
for the $6j$ symbol given in \eqref{PRasymt} of Appendix A.1,
we realize that it represents the semiclassical functional, namely
the semiclassical limit of a path--sum over all quantum fluctuations, to be associated with  
the simplest $3$--dimensional "spacetime", an Euclidean tetrahedron $T$. In fact the argument
in the exponential reproduces the  Regge action  $S^3(\ell )$ for $T$ since in the present case
$(D-2)$ simplices are
$1$--dimensional (edges) and $\text{Vol}^{(D-2)}(\sigma_i)$ in (\ref{ReAction})
are looked at as  the associated  edge lengths, see the introductory part of Appendix A.\\ 
More in general, we denote by
$\mathcal{T}^3\,(j)\rightarrow \mathcal{M}^3$
a particular triangulation of a closed $3$--dimensional Regge manifold 
$\mathcal{M}^3$ (of fixed topology) obtained by assigning $SU(2)$ spin variables
$\{j\}$ to the edges of $\mathcal{T}^3$. The assignment must satisfy a number of conditions,
better illustrated if we introduce the {\em state functional}
associated with $\mathcal{T}^3 (j)$, namely
\begin{equation}\label{PRstfunct}
\mathbf{Z}[\mathcal{T}^3(j) \rightarrow \mathcal{M}^3; L]=
\Lambda(L)^{-N_0}\prod_{A=1}^{N_1} (-1)^{2j_A} \mathsf{w}_A\prod_{B=1}^{N_3}
\phi_B
\begin{Bmatrix}
j_1 & j_2 & j_3 \\
j_4 & j_5 & j_6
\end{Bmatrix}_B
\end{equation}
where $N_0, \, N_1,\, N_3$ are the number of vertices, edges and tetrahedra 
in $\mathcal{T}^3(j)$, $\Lambda (L)=4L^3/3C$ ($L$ is a fixed length and $C$ an arbitrary constant),
$\mathsf{w}_A \doteq$ $(2j_A+1)$ are the dimensions of 
irreducible representations of $SU(2)$ which weigh the edges,
$\phi_B =$ $(-1)^{\sum_{p=1}^6 j_p}$ and $\{:::\}_B$ 
are $6j$ symbols to be associated with the tetrahedra of the triangulation. 
Finally, the Ponzano--Regge 
{\em state sum} is obtained by summing over triangulations
corresponding to all assignments of spin variables $\{j\}$ bounded by the cut--off $L$
\begin{equation}\label{PRstsum}
\mathbf{Z}_{PR}\,[\mathcal{M}^3]\;=\;
\lim_{L\rightarrow \infty}\:
\sum_{\{j\}\leq L}
\mathbf{Z}\; [\,\mathcal{T}^3(j) \rightarrow \mathcal{M}^3; L\,]\;,
\end{equation} 
where the cut--off is formally removed by taking the limit in
front of the sum.\\

It is not easy to review in short the huge number of implications and further improvements 
of Ponzano--Regge state sum functional (\ref{PRstsum}), as well as its deep and somehow surprising 
relationships with so many different issues in modern theoretical physics and in pure mathematics.
We are going to present in the rest of this section a limited number of items, whose selection is made
mainly on the basis of their relevance for (quantum) computational problems raised in the next section
(we remind however the importance of this model in the so--called "loop" approach to 
quantum gravity \cite{Rov}, see also \cite{ReWi}).
 
\begin{itemize}
\item[(a)] 
As already noted in \cite{PoRe}, the state sum $\mathbf{Z}_{PR}\,[\mathcal{M}^3]$ is a topological
invariant of the manifold $\mathcal{M}^3$, owing to the fact that its value is actually independent  of
the particular triangulation, namely does not change under
suitable combinatorial transformations. Remarkably,  
these "moves" are expressed algebraically in terms of the
relations given in Appendix A.2, namely the Biedenharn-Elliott identity  
\eqref{BEid}
--representing the moves 
(2 {\em tetrahedra}) $\leftrightarrow$  (3 {\em tetrahedra})-- and of both the 
Biedenharn--Elliott identity and the orthogonality conditions 
\eqref{ort6j}
for $6j$ symbols, which represent the barycentric move together its inverse, namely 
(1 {\em tetrahedra}) $\leftrightarrow$  (4 {\em tetrahedra}). 
\item[(b)] In \cite{TuVi} a "regularized" version of \eqref{PRstsum} --based on representation
theory of a quantum deformation 
of the group $SU(2)$--  was proposed and shown to be a well--defined 
{\em quantum invariant} for closed 3--manifolds\footnote{
The adjective "quantum" refers here to "deformations" of
semi--simple  Lie groups introduced by the Russian School of theoretical physics
in the 1980's in connection with inverse scattering theory. From
the mathematical viewpoint the Turaev--Viro invariant, unlike the Ponzano--Regge
state sum functional, is always finite and has been evaluated explicitly
for some classes of $3$--manifolds.}.\\
 Its expression reads
\begin{equation}\label{TVstsum}
\mathbf{Z}_{\,TV}\,[\mathcal{M}^3;q]\,=\,\sum_{\{j\}}\;\mathbf{w}^{-N_0}\,
\prod_{A=1}^{N_1} \mathbf{w}_A
\,\prod_{B=1}^{N_3} \;
\begin{vmatrix}
j_1 & j_2 & j_3 \\
j_4 & j_5 & j_6
\end{vmatrix}_B \,,  
\end{equation}
\noindent 
where the summation is over all  $\{j\}$ labeling highest weight irreducible representations of $SU(2)_q$ 
($q=\exp\{2\pi i /r\}$, with $\{j=0,1/2,1 \dots, r-1\}$), $\mathbf{w}_A\doteq$ 
$(-1)^{2j_A}[2j_A+1]_q$ 
where $[\,]_q$ denote a  quantum integer, 
$\mathbf{w}=2r/(q-q^{-1})^2$ and $|:::|_B$ represents here the q--$6j$ symbol whose entries are the 
angular momenta $j_{i}, i =1,\dots ,6$ associated with tetrahedron $B$.
If the deformation parameter q is set to $1$ one gets
$\mathbf{Z}_{\,TV}\,[\mathcal{M}^3;1]$ $=\mathbf{Z}_{PR}\,[\mathcal{M}^3]$.\\
It is worth noting that
the q--Racah polynomial --associated with the q--$6j$ by a procedure that matches 
with what can be done in the $SU(2)$ case, see (\ref{F43}) in Appendix A.2--
stands at the top of Askey's q--hierarchy collecting
 orthogonal q--polynomials of one discrete or continuous variable. On the other hand,
the discovery of the Turaev--Viro invariant has provided major developments in
the branch of mathematics known as  geometric topology \cite{Oht}.
\item[(c)] The Turaev--Viro or Ponzano--Regge state sums as defined above can be generalized in 
many directions. For instance, they can be extended to simplicial $3$--manifold endowed with
a $2$--dimensional boundary \cite{CaCaMa1} and to $D$--manifolds \cite{CaCaMa2}
(giving rise to topological invariants related to suitable (discretized) 
topological quantum field theory of the Schwarz type \cite{Kau}).
\item[(d)] 
The fact that the Turaev--Viro state sum is a topological invariant of
the underlying (closed) $3$--manifold
reflects a crucial physical property of gravity in dimension $3$ which
makes it different from the corresponding $D=4$ case. Loosely speaking,
the gravitational field does not possess local degrees of freedom
in $D=3$, and
thus  any "quantized"  functional can depend only on global features
of the manifold encoded into its overall topology.
Actually the invariant \eqref{TVstsum} can be  shown to be equal to the 
square of the modulus of the Witten--Reshetikhin--Turaev invariant, which in turn represents
a quantum path--integral of an $SU(2)$ Chern--Simons topological field theory
--whose classical action can be shown to be equivalent
to Einstein--Hilbert action \cite{Car}--  
 written for a closed oriented manifold
$\mathcal{M}^3$ \cite{Wit1,ReTu}. Then there exists
a corresponence
\begin{equation}\label{TVCS}
\mathbf{Z}_{\,TV}\,[\mathcal{M}^3;q\,]\,\longleftrightarrow\,
|\,\mathbf{Z}_{\,WRT}\,[\mathcal{M}^3;k\,]\,|^2\,,
\end{equation}
where the "level" $k$ of the Chern--Simons functional is related to the
deformation parameter $q$ of the quantum group.
\end{itemize}
Despite the "topological" nature of Turaev--Viro (Ponzano--Regge) state sum and
Witten--Reshetikhin--Turaev functionals in case of closed $3$--manifolds,
whenever a $2D$--dimensional boundary occurs in $\mathcal{M}^3$,
giving rise to a pair $(\mathcal{M}^3, \Sigma)$, where $\Sigma$ is an
oriented surface (or possibly the disjoint union of a finite number of surfaces),
things change radically. For instance, if we add a boundary to the manifold in 
Witten--Reshetikhin--Turaev quantum functional, the theory induced on
$\Sigma$ is a Wess--Zumino--Witten (WZW)--type Conformal Field Theory (CFT)
\cite{Car}, endowed with non--trivial quantum degrees of
freedom. In particular, the frameworks outlined above
can be exploit to establish a direct correspondence
between $2D$ Regge triangulations and punctured Riemann surfaces, thus
providing a novel characterization of the WZW model on triangulated
surfaces on any genus \cite{ArCaDa} at a fixed level $k$.\\ 
We cannot enter here into many  technical details
on these developments. It should be sufficient to remark
that, when addressing "boundary" CFT, the geometric role  of the  quantum 
tetrahedron shades out, while its algebraic
content is enhanced given that the (q)--$6j$--symbol plays the role of a "duality" (or "fusion")
 matrix, similar to a "recoupling coefficient" between different
basis sets, as (\ref{6j1}) in Appendix A suggests.

\begin{itemize}
\item[(e)] In \cite{M2Ra} a $(2+1)$--dimensional  decomposition of Euclidean gravity
(which takes into account the correspondence (\ref{TVCS})) is shown to be equivalent,
under mild topological assumptions, to a Gaussian $2D$ fermionic system, whose partition 
function takes into account the underlying $3D$ topology. More precisely, 
the partition function for free fermions propagating along "knotted loops" inside
a $3$--dimensional sphere corresponds to a $3D$ Ising model  on so--called
knot--graph lattices. On the other hand, the formal expression of $3D$ Ising
partition function for a dimer covering of the underlying graph lattice
can be shown to
coincide with the permanent of the generalized incidence matrix of the
lattice \cite{CeRaZe,ReZe}. 
Recall first that the permanent of an $n \times n$
matrix $A$ is given by
\begin{equation}\label{per}
\text{per} [A]\,=\, \sum_{\sigma \in \mathsf{S}_n}\;
\prod_{i=1}^{n}\; a_{i, \sigma(i)}
\end{equation}
where $a_{i, \sigma(i)}$ are minors of the matrix,
$\sigma(i)$ is a permutation of the index $i=1,2,\dots,n$
and $\mathsf{S}_n$ is the symmetric group on $n$ elements.
A graph lattice ${\mathfrak{G}}$ associated with a fixed orientable surfaces
{$\Sigma$} of genus $g$ embedded in  $S^3$ may be 
constructed by resorting to the so--called "surgery link" presentation.
Then the incidence matrix of
such piecewise linear graph with, say, $\mathfrak{n}$
vertices, is defined as an $\mathfrak{n} \times \mathfrak{n}$ 
matrix $A= (a_{ij})$ with entries in $(1,0)$ according to whether
vertices $i,j$ are connected by an edge or not. Finally,
the Ising partition function turns out to be a weighted sum --over all
possible configurations of knot--graph lattices-- of 
suitable "determinants" of generalized forms
of the incidence matrices which take into account the topology
of the underlying manifold. We skip however other technical
details and refer to \cite{MaRa2} for a short account of these
results (which will be briefly reconsidered in the
following section in the context of quantum computational questions).
\end{itemize}

The deep relationship between $3D$ quantum field theories
that share a "topological" nature
and (solvable) lattice models in $2D$,
sketched in the last item by resorting to a specific example,
 was indeed predicted in the pioneering paper by E. Witten \cite{Wit2}.
 Not so surprisingly, the basic quantum functional
that realizes this connection was identified there with
{\em the expectation value of a  certain tetrahedral  configuration
of braided Wilson lines}, where  "Wilson lines" are quantum observables 
associated with "particle trajectories" that in general
look like  sheafs of braided strands propagating 
from a surface $\Sigma_1$ to another $\Sigma_2$, both embedded in a $3D$
background.

\section{6j symbol and quantum algorithms}

The model for  universal quantum computation proposed in \cite{MaRa1},
the "spin network" simulator, 
is based on the (re)coupling theory of $SU(2)$ angular momenta
as formulated in the basic texts \cite{BiLo8,BiLo9} on the quantum
theory of angular momentum and the Racah--Wigner algebra respectively.
At the first glance the spin network simulator can
be thought of as a non--Boolean generalization of the Boolean
{\em quantum circuit model} \footnote{
Recall that this scheme
is the quantum version of the classical Boolean circuit in which strings of the basic
binary alphabet  $(0,1)$ are replaced by
collections of "qubits", namely quantum states in $(\mathbb{C}^{2})^{\otimes N}$,
and the gates are unitary transformations that can be expressed, similarly to what happens in the
 classical case, as suitable sequences of "elementary" gates associated with the Boolean logic operations
{\em and, or, not}.} \cite{NiCh}, 
 with finite--dimensional, binary coupled computational
Hilbert spaces associated with $N$ mutually commuting angular momentum
operators and unitary gates expressed in terms of:\\
i) recoupling coefficients ($3nj$ symbols) between inequivalent binary
coupling schemes of $N=(n+1)$ $SU(2)$--angular momentum variables ($j$--gates);\\
ii) Wigner rotations in the eigenspace of the total angular momentum $\mathbf{J}$ ($M$--gates)
(that however will not be taken into account in what follows, see section 3.2
of \cite{MaRa1} for details) 

\bigskip\bigskip  
\begin{center} 
\thicklines  
\begin{picture}(250,280)(-94,-135)  
\put(-140,37){\framebox(130,32){BOOLEAN Q-CIRCUIT}}  
\put(-140,-32){\framebox(130,32){TOPOLOGICAL QFT}}  
\put(-78,28){\vector(0,-1){25}}
\put(-77,3){\vector(0,1){25}}
\put(-77,28){\vector(0,-1){25}}
\put(-78,3){\vector(0,1){25}}
\put(108,87){\vector(0,1){25}}
\put(109,87){\vector(0,1){25}}
\put(108,29){\vector(0,-1){25}}
\put(109,4){\vector(0,1){25}}
\put(109,29){\vector(0,-1){25}}
\put(108,4){\vector(0,1){25}}
\put(108,-51){\vector(0,-1){25}}
\put(109,-76){\vector(0,1){25}}
\put(109,-51){\vector(0,-1){25}}
\put(108,-76){\vector(0,1){25}}
\put(0,51){\vector(1,0){40}} 
\put(0,50){\vector(1,0){40}} 
\put(35,-105){\vector(-3,4){49}}
\put(35,-106){\vector(-3,4){49}}
\put(55,116){\framebox(120,46){Q-AUTOMATA}} 
\put(55,37){\framebox(120,46){}} 
\put(115,70){\makebox(0,0){GENERALIZED}}
\put(115,50){\makebox(0,0){Q-CIRCUIT}}
\put(55,-46){\framebox(120,46){}}
\put(53,-48){\framebox(124,50){}}
\put(67,-16){\makebox{SPIN NETWORK}} 
\put(67,-37){\makebox{Q-SIMULATOR}}
\put(55,-125){\framebox(120,46){}} 
\put(115,-92){\makebox(0,0){STATE SUM}} 
\put(115,-112){\makebox(0,0){MODELS}} 
\end{picture}  
\end{center} 
\bigskip 

In the diagram we try to summarize various aspects of
the spin network simulator together with its
relationships with other models for Q--computation, in the
light of underlying physical frameworks discussed in the previous section.\\
On the left--hand portion of the diagram
the standard Boolean quantum
circuit is connected with a double arrow to the so--called topological approach to quantum computing
developed in \cite{FrLaWa} (based, by the way,  on the
Witten--Reshetikhin--Turaev approach quoted
in item (d) of the previous section). 
This means in practice that these two models of computation can be 
efficiently "converted"  one into the other.
The Boolean case is connected one--way to the
box of the generalized Q--circuit  because it is actually
a particular case of the latter  
when all $N$ angular momenta are $\frac{1}{2}$--spins.\\
On the right--hand column, the double arrows stemming from
the box of the spin network Q--simulator relate it to its reference models: from the
viewpoint of quantum information theory
it is a generalized Q--circuit, as already noted before,
while its physical setting can be
assimilated to state sum--type models discussed in the first part of the previous section.\\
 The upper arrow is to be meant as generating, from the
general Q--computational scheme, families of "finite--states" Q--automata able to 
process in an efficient way a number of specific algorithmic problems
that on a classical computer would require an exponential amount
of resources ({\em cfr.} the end of this section).\\

Besides the features described above,
the kinematical structure of the Q--spin network  
complies with all the requisites of an universal Q--simulator as defined by 
Feynman in \cite{Fey}, namely\\
$\bullet$  {\em locality}, reflected in the binary bracketing structure of the computational 
Hilbert spaces, which  bears on 
the existence of poly--local, two--body  interactions;\\
$\bullet$ {\em discreteness of the computational space}, reflected in the combinatorial
structure of the (re)coupling theory of $SU(2)$ angular momenta \cite{BiLo9,Russi,YuLeVa};\\
$\bullet$ {\em discreteness of time}, given by the possibility of selecting controlled,  step--by--step 
applications of sequences  of unitary operations for the generation of (any) process of computation;\\ 
$\bullet$ {\em universality}, guaranteed 
by the property that any unitary transformation 
operating on binary coupled Hilbert spaces  (given by $SU(2)$
 $3nj$ symbols) can be reconstructed by taking a finite 
sequence of  Racah--Wigner transforms implemented by expression
of the type given in (\ref{6j1}) of Appendix A (possibly apart from phases factors),
as shown in \cite{BiLo9}, topic 12.

Then the Wigner $6j$ symbol plays a prominent role also in the spin network Q-simulator
scheme, where it is the  "elementary" unitary operation, from which any
"algorithmic" procedure can be built up. The meaning of the identities (\ref{BEid})
(\ref{ort6j}) satisfied by the $6j$'s in the present context is analyzed at length in \cite{MaRa1},
(section 4.2 and Appendix A) and  can be related to the notion of intrinsic "parallelism"
of quantum computers.

A caveat is however in order: the complexity class of any classical [quantum] algorithm is defined
with respect to a "standard" classical [quantum] model of computation\footnote{Recall that
a quantum algorithm for solving a given computational  problem is "efficient" if it belongs to the
complexity class $\mathbf{BQP}$, namely the class of problems that can be
solved in polynomial time by a Boolean  Q--circuit with a fixed bounded error in terms
of the "size" of a typical input. In most examples the size of the input
is measured by the length of the string of qubits necessary to
encode the generic sample of the algorithmic problem, as happens with the binary 
representation of
an integer number in calculations aimed to factorize it in prime factors.}. 
At the quantum level,
such a reference model is the Boolean Q--circuit \cite{NiCh}, and thus what is necessary to verify is that
a  $6j$ symbol with generic entries can be efficiently (polynomially) processed by a 
suitably designed Q--circuit. Note first that a $6j$ symbol with fixed
entries, due to the finiteness of the Racah sum rule (see  (\ref{F43}) in Appendix A.2),
can be efficiently computed classically. On the other hand,
the $6j$ is a $(2d+1) \times (2d+1)$ unitary matrix representing
a change of basis, as given explicitly in (\ref{6j1}) of Appendix A, with $j_{12}, 
j_{23}$ representing matrix indices running over an  interval of length $2d+1$
in integer steps. 
Thus the evaluation of the complexity class  of this problem consists is asking whether,
as $d$ increases, the calculation of the $6j$ falls into the $\mathbf{BQP}$ class.
The circuit which implements such task has been designed in
\cite{GaMaRa1} for the case of the $SU(2)_q$ $6j$ for each $q$ = root of unity,
while the analog problem involving
the "classical", $SU(2)$ $6j$ is still open.\\

In the last few years  two of the authors, in collaboration with S. Garnerone,
have developed, on the basis of the spin network
simulator setting \cite{MaRa1}, a new approach to deal with
classes of algorithmic problems that classically admit only
exponential time algorithms. The problems in questions arise
in the physical context of $3D$ topological
quantum field theories discussed in the previous section
in the light of the fundamental result relating a topological invariant of knots, the
Jones polynomial \cite{Jon}, with a quantum observable given by the vacuum expectation value
of
a Wilson "loop" operator  \cite{Wit3} associated with closed knotted curves
in the Witten--Reshetikhin--Turaev background model.\\ 
Without entering into technical details, efficient (polynomial time)
quantum algorithms for approximating (with an error that can be made as small
as desired) generalizations of Jones polynomial have been found in
 \cite{GaMaRa1,GaMaRa2}, while the case of topological invariants
of $3$--manifolds has been addressed in \cite{GaMaRa3}.
The relevance  in having solved this kind of problems stems from the fact that an 
approximation of the Jones polynomial is sufficient 
to simulate any polynomial quantum computation \cite{BoFrLo}.

Summing up, the construction of such quantum algorithms actually 
bears on the interplay of three different contexts 
\begin{enumerate}
 \item a topological context, where the problem is well--posed 
 and makes it possible to recast the initial instance 
 from the topological language of knot theory to the algebraic language of braid group theory,
as reviewed in \cite{GaMaRa4};
 \item a field theoretic context, where tools from $3D$ topological quantum field
 and associated $2D$ conformal field theory are used to provide 
 a unitary representation of the braid group;
\item a quantum information context, where the basic features 
of quantum computation are used to efficiently solve the 
original problem formulated in a field theoretic language.
\end{enumerate}
In the light of   remark (e) at the end of section 2, further analysis 
of relationships between specific $3D$ topological quantum field theories
and (solvable) lattice models in $2D$ in the quantum--computational context
would represent a major improvement not only from a theoretical viewpoint,
but also  in view of possible physical implementations.
In \cite{MaRa2} some preliminary progress has been achieved for
establishing a quantum algorithm for the evaluation of the permanent 
(\ref{per}) associated with the partition function of the Ising model on knot--graph lattices.
As shown in \cite{Loe} by resorting to numerical simulations, such a 
computational problem can be related to the computation of
Jones invariants on suitably defined configurations, thus providing 
further evidence of the "universality" of any one of the quantum algorithms
quoted above.

In conclusion, we hope to have been able to illustrate in sufficient details
the role of the Wigner $6j$ symbol (or the q--$6j$) as an universal 
building block unifying such different fields as  quantum geometry,
topological quantum field theory, statistical lattice models and quantum computing.

The interplay between solvability and computability 
within the framework of quantum Witten--Reshetikhin--Turaev  theory and solvable
lattice models deserves however a few more  comments.
Unlike perturbatively renormalizable quantum field theory  
--which represent the basic tool in the standard model in particle physics, 
where the  physically measurable quantities are obtained as finite limits of
infinite series in the physical coupling constant-- quantum WRT theory is actually "solvable" 
since functionals of type (\ref{TVCS}) and   (\ref{TVstsum}),
as well as  Wilson loop observables, are  sums of a {\em finite number}
of terms for each fixed value of the deformation parameter q.
Actually such finiteness property reflects  the existence of
a deeper  algebraic symmetry stemming from braid group representations
and associated Yang--Baxter equation, see {\em e.g.} \cite{Wit3,GaMaRa4} and references therein\footnote{
This notion of solvability might be viewed as the quantum analog of the
property of "complete integrability" in classical mechanics.
Recall that  integrable  systems admit a sufficient 
number of conserved quantities that make it possible to solve
explicitly Newton equations of motion. These "constants of motions" are endowed with 
a suitable algebraic structure under Poisson bracketing which 
is related in turn to complete integrability
owing to Arnold--Liouville theorem.}.
The issue of  computability of all the relevant quantities
of quantum WRT theory, and in particular of the Jones polynomial, 
is ultimately  related to
solvability/finiteness of the underlying theory.
Thus the existence of "efficient"  computational protocols should
help in sheding light  on the open question concerning the validation of
the heuristic procedure associated with the path--sum
quantization scheme (may be also in other contexts).
Turning the argument upside down, the search for new efficient quantum algorithms
for processing "invariant quantities" characterizing suitably decorated
lattice, graphs, surfaces, {\em etc.} represents an original
and possibly very fruitful approach for understanding
the underlying physical models with
respect to their (yet unknown) integrability properties.

 \section*{Appendix A: the Wigner 6j symbol and its symmetries}
 
Given three angular momentum operators
${\bf J}_1,{\bf J}_2, {\bf J}_3$  --associated with three
kinematically independent quantum systems-- the
Wigner--coupled Hilbert space of the composite system 
is an eigenstate of the total angular momentum
\begin{equation}\label{jtot}
{\bf J}_1\,+\,{\bf J}_2\,+\,{\bf J}_3\;\doteq\;{\bf J}
\end{equation}
\noindent and of its projection $J_{z}$ along the quantization axis. 
The degeneracy can be completely removed by considering
binary coupling schemes such as 
$({\bf J}_1\,+\,{\bf J}_2)\,+\,{\bf J}_3$ and
${\bf J}_1\,+\,({\bf J}_2\,+\,{\bf J}_3)$, and by introducing
intermediate angular momentum operators defined by
\begin{equation}\label{j12}
({\bf J}_1\,+\,{\bf J}_2)  = {\bf J}_{12};\;\,
{\bf J}_{12}\,+\,{\bf J}_3  = {\bf J}
\end{equation}
and
\begin{equation}\label{j23}
({\bf J}_2\,+\,{\bf J}_3)  = {\bf J}_{23};\;\; 
{\bf J_1}\,+\,{\bf J}_{23} = {\bf J},
\end{equation}
respectively. In Dirac notation the simultaneous
eigenspaces of the two complete sets of commuting operators
are spanned by basis vectors
\begin{equation}\label{basis}
|j_1 j_2 j_{12} j_3; \,j m\rangle\;\; \text{and}\;\;
|j_1 j_2 j_3 j_{23};\, j m\rangle,
\end{equation}
where  $j_1, j_2, j_3$ denote eigenvalues of the corresponding operators,
$j$ is the eigenvalue of   ${\bf J}$
and $m$ is the total magnetic quantum number with range
$-j \leq m \leq j$ in integer steps. 
Note that 
$j_1, j_2, j_3$ run over $\{0,\tfrac{1}{2}, 1, \tfrac{3}{2}, 2, \dots \}$
(labels of $SU(2)$ irreducible representations), 
while 
 $|j_1-j_2|\leq j_{12}\leq j_1+j_2$ and 
  $|j_2-j_3|\leq j_{23}\leq j_2+j_3$ (all quantum numbers are in $\hbar$
  units).
  
  The  Wigner $6j$ symbol expresses the transformation between the two
  schemes (\ref{j12})
and (\ref{j23}), namely  
 \begin{equation}\label{6j1}
|j_1 j_2 j_{12} j_3; \,j m \rangle
= \sum_{j_{23}}\,
[(2j_{12}+1) (2j_{23}+1)]^{1/2}\,
\begin{Bmatrix}
j_1 & j_2 & j_{12}\\
j_3 & j & j_{23}
\end{Bmatrix}
|j_1 j_2 j_3 j_{23};\, j m \rangle
  \end{equation}
 apart from a phase factor\footnote{
Actually this expression should contain the Racah W--coefficient
$W(j_1 j_2 j_3 j;j_{12} j_{23})$ which differs from the $6j$
by the factor $(-)^{j_1 + j_2 + j_3 + j}$. Recall that
$(2j_{12}+1)$ and  $(2j_{23}+1)$ are the dimensions of the
representations labeled by $j_{12}$ and  $j_{23}$, respectively.}.
It follows that the quantum mechanical probability
\begin{equation}\label{Pr6j}
P\,=\,[(2 j_{12}+1) (2 j_{23}+1)]\,
\begin{Bmatrix}
j_1 & j_2 & j_{12}\\
j_3 & j & j_{23}
\end{Bmatrix}^2 
\end{equation}
represents the probability that a system prepared in a state of the
coupling scheme (\ref{j12}), where $j_1,  j_2 , j_3, j_{12}, j$
have definite magnitudes, will be measured to be in a state 
of the coupling scheme (\ref{j23}).

The $6j$ symbol may be written as sums of products of four Clebsch--Gordan
coefficients or their symmetric counterparts, the Wigner $3j$ symbols. The relations between  $6j$
and  $3j$ symbols are given explicitly by (see {\em e.g.} \cite{Russi})
\begin{equation}\label{6j2}
\begin{Bmatrix}
a & b & c\\
d & e & f
\end{Bmatrix}=
\sum (-)^{\Phi}
\begin{pmatrix}
a & b & c\\
\alpha & \beta & -\gamma
\end{pmatrix}
\begin{pmatrix}
a & e & f\\
\alpha & \epsilon& -\varphi
\end{pmatrix}
\begin{pmatrix}
d & b & f\\
-\delta & \beta & \varphi
\end{pmatrix}
\begin{pmatrix}
d & e & c\\
\delta & -\epsilon & \gamma
\end{pmatrix}
\end{equation}
where $\Phi= d+e+f+ \delta + \epsilon + \varphi$.
Here Latin letters stand for $j$--type labels
(integer or half--integers non--negative numbers)
while Greek
letters denote the associated magnetic quantum numbers
(each varying in integer steps between $-j$ and $j$,
$j \in \{a,b,c,d,e,f\}$). The sum is over all possible values of
$\alpha, \beta, \gamma,$ $\delta, \epsilon, \varphi$ 
with only three summation indices being independent.\\
On the basis of  the above decomposition it can be shown
that the $6j$ symbol is invariant under any permutation
of its columns or under interchange the upper and lower arguments in each
 of any two columns. These algebraic relations involve $3! \times 4 =24$
 different $6j$ with the same value and are referred to as 
 {\em classical symmetries} as opposite to "Regge" 
symmetries to be discussed in A.2.

The $6j$ symbol is naturally endowed with a
geometric symmetry, the {\em tetrahedral
symmetry}, as  the reproduction in
Fig. 1 suggests. Note first that
each $3j$ (or Clebsch--Gordan) coefficient vanishes unless
its $j$--type entries satisfy the triangular condition,
namely $|b-c|\leq a\leq b+c$, {\em etc.}. 
This suggests that each of the four $3j$'s in (\ref{6j2})
can be be associated with either a $3$--valent vertex
or a triangle. Accordingly,
there are two graphical representation of
the $6j$ exhibiting its symmetry properties. Here we adopt
the three--dimensional picture introduced in the seminal paper
by Ponzano and Regge 
\cite{PoRe}, rather than Yutsis' "dual"
representation as a complete graph on four vertices 
\cite{YuLeVa}.
Then the $6j$ is thought of as a real solid tetrahedron $T$
with edge lengths $\ell_1=a + \tfrac{1}{2},
\ell_2=b + \tfrac{1}{2},$ $ \dots, \ell_6=f+ \tfrac{1}{2}$
in $\hbar$ units\footnote{
The $\tfrac{1}{2}$--shift is shown to be crucial
in the analysis developed in 
\cite{PoRe}:
for high quantum numbers the length $[j(j+1)]^{1/2}$ of an angular
momentum vector is closer to $j+ \tfrac{1}{2}$ in the semiclassical limit.} 
and triangular faces associated with the triads
$(abc)$, $(aef)$, $(dbf)$, $(dec)$. This implies in particular
that the quantities $q_1=a+b+c$, $q_2=a+e+f$,
$q_3=b+d+f$, $q_4=c+d+e$ (sums of the edge lengths of each face),
$p_1=a+b+d+e$, $p_2=a+c+d+f$,
$p_3=b+c+e+f$ are all integer with $p_h \geq q_k$ ($h=1,2,3$, $k=1,2,3,4$).
The conditions addressed so far are in general sufficient to guarantee the existence
of a non--vanishing $6j$ symbol, but they are not enough to ensure the existence of 
a geometric  tetrahedron $T$ living in Euclidean $3$--space with the given edges. 
More precisely,
$T$ exists in this sense if ({\em and only if}, see the discussion in the introduction of
\cite{PoRe}) its square volume $V(T)^2 \equiv V^2$, evaluated by means of
the Cayley--Menger determinant, is positive.

The features of the "quantum tetrahedron" outlined above represent the foundations
of a variety of  results, some of which were discovered in the golden age 
of quantum mechanics 
and have been widely used  in old and present applications to atomic and molecular physics. 
In this paper we have tried to convey 
at least a few applications of this intriguing object in modern theoretical physics,
while in the rest of this appendix we are going to 
complete the mathematical background needed 
in the previous sections, focusing  in particular  on  semiclassical
analysis and results from special function theory.

\subsection*{A.1 Ponzano--Regge asymptotic formula}

The Ponzano--Regge asymptotic formula for the $6j$
symbol reads \cite{PoRe}
\begin{equation}\label{PRasymt}
\begin{Bmatrix}
a & b & d\\
c & f & e
\end{Bmatrix}
\;\sim\;\; \frac{1}{\sqrt{24 \pi V}}\;
\exp\,\left\{i\,\left(\sum_{r=1}^{6}\,\ell_r \, \theta_r \,+\,\frac{\pi}{4}
\right)\right\}
\end{equation}

\noindent where the limit is taken for all entries $\gg 1$ (recall that $\hbar =1$)
and $\ell_r \equiv j_r +1/2$
with $\{j_r\}=\{a,b,c,d,e,f\}$. 
$V$ is the Euclidean volume of the tetrahedron $T$ 
and $\theta_r$ is the angle between the outer normals to the faces which
share the edge $\ell_r$.\\
From a quantum mechanical  viewpoint, the above probability amplitude
has the form of a semiclassical (wave) function since the factor 
$1/\sqrt{24 \pi V}$ is slowly varying with respect to the spin variables while
the exponential is a rapidly oscillating dynamical phase. 
Such kind of asymptotic behavior complies  with
Wigner's semiclassical estimate for the probability, namely  
$\left\{\begin{smallmatrix}
a & b & d\\
c & f & e
\end{smallmatrix}\right\}^{\,2}\sim 1/12 \pi\,V\,$,
to be compared with the quantum probability
given in (\ref{Pr6j}).
Moreover, according to Feynman path sum
interpretation of quantum mechanics \cite{FeHi}, the argument of the exponential 
in \eqref{PRasymt} must represent a classical action, and indeed it can be read as
$\sum \mathsf{p}\,\dot{\mathsf{q}}$ for pairs $(\mathsf{p},\mathsf{q})$ 
of canonical variables
(angular momenta and conjugate angles). 
Such an interpretation has been improved recently by resorting to
multidimensional WKB theory for integrable systems
and geometric quantization methods \cite{AqHaLi}.

\subsection*{A.2 Racah hypergeometric polynomial}
 
The generalized hypergeometric series, denoted by
$_pF_q$, is defined on $p$ real or complex numerator parameters
$a_1,a_2,\dots,a_p$, $q$ real or complex denominator parameters 
 $b_1,b_2,\dots,b_q$ and a single variable $z$ by 
\begin{equation}\label{Fpq}
_pF_q 
\begin{pmatrix}
a_1 & \dots & a_p & \, & \, \\
\, & \, & \, & ; & z \\
b_1 & \dots & b_q & \, & \,
\end{pmatrix}
\,=\,
\sum_{n=0}^{\infty}\,
\frac{(a_1)_n \cdots  (a_p)_n}{(b_1)_n \cdots  (b_p)_n}\;
\frac{z^n}{n!}\,,
\end{equation}
where $(a)_n=a(a+1)(a+2)\cdots(a+n-1)$ denotes a rising factorial
with $(a)_0=1$. If one of the numerator parameter is a negative integer,
as actually happens in the following formula,
the series terminates and the function is a polynomial in $z$.\\
The key expression for relating the $6j$ symbol to hypergeometric
functions is given by the well--known Racah sum rule (see {\em e.g.}
\cite{BiLo9}, topic 11 and \cite{Russi}, Ch. 9 also for the original references).
The final form of the so--called {\em
Racah polynomial} is written in terms of the $_4F_3$ hypergeometric function
evaluated at $z=1$ according to
\begin{equation*}
\begin{Bmatrix}
a & b & d\\
c & f & e
\end{Bmatrix}
\,=\,\Delta(abe)\,\Delta(cde)\,\Delta(acf)\,\Delta(bdf) \;(-)^{\beta_1} (\beta_1+1)!
\end{equation*}  
\begin{equation}\label{F43}
\times\,\frac{_4F_3
\left(
\begin{smallmatrix}
\alpha_1-\beta_1 & \alpha_2-\beta_1 & \alpha_3-\beta_1 & \alpha_4-\beta_1 & \, & \,\\
\, & \, & \, &  &; & 1 \\
-\beta_1-1\, & \beta_2-\beta_1+1 \, &  \beta_3-\beta_1+1 & \, & \, & \,
\end{smallmatrix}\right)}
{(\beta_2-\beta_1)! (\beta_3-\beta_1)! (\beta_1-\alpha_1)!
(\beta_1-\alpha_2)! (\beta_1-\beta_3)! (\beta_1-\alpha_4)!}\;,
\end{equation}
where 
$$
\beta_1=\, \min (a+b+c+d; a+d+e+f; b+c+e+f)
$$
and the parameters $\beta_2,\beta_3$
are identified in either way with the
pair remaining in the $3$--tuple
$(a+b+c+d; a+d+e+f; b+c+e+f)$
after deleting $\beta_1$. The four $\alpha$'s 
may be identified with any permutation of
 $(a+b+e; c+d+e;$ $a+c+f; b+d+f)$. Finally, the 
$\Delta$--factors in front of $_4F_3$
 are defined, for any triad $(abc)$ as
 $$
 \Delta\, (abc)\,=\,
 \left[
 \frac{(a+b-c)!(a-b+c)! (-a+b+c)!}{(a+b+c+1)!}
 \right]^{1/2}
 $$
 Such a seemly complicated notation is indeed
 the most convenient for the purpose of listing
further interesting properties of the Wigner $6j$ symbol.
\begin{itemize}
  \item The Racah polynomial is placed at the top of the
  Askey hierarchy including all of hypergeometric orthogonal polynomials
  of one (discrete or continuous) variable \cite{Askey}.
  Most commonly encountered families of special functions in  quantum mechnics
 are obtained from the Racah polynomial by applying
 suitable limiting procedures, as recently
reviewed in \cite{Ragni}. Such an unified scheme
provides in a straightforward way the algebraic {\em defining relations}
of the Wigner $6j$ symbol viewed as an orthogonal polynomial
of one discrete variable, {\em cfr.} (\ref{F43}). By resorting
to standard notation from the quantum theory of angular momentum,
the defining relations are:\\
the Biedenharn--Elliott identity ($R=a+b+c+d+e+f+p+q+r$):
\begin{align}\label{BEid}
\sum_{x}(-)^{R+x}\,(2x+1)&\begin{Bmatrix}
a & b & x\\
c & d & p
\end{Bmatrix}
\begin{Bmatrix}
c & d & x\\
e & f & q
\end{Bmatrix}
\begin{Bmatrix}
e & f & x\\
b & a & r
\end{Bmatrix}\nonumber\\
& =\;
\begin{Bmatrix}
p & q & r\\
e & a & d
\end{Bmatrix}
\begin{Bmatrix}
p & q & r\\
f & b & c
\end{Bmatrix};
\end{align}
the orthogonality relation ($\delta$ is the Kronecker delta)
\begin{equation}\label{ort6j}
\sum_{x}\,(2x+1)\,
\begin{Bmatrix}
a & b & x\\
c & d & p
\end{Bmatrix}
\begin{Bmatrix}
c & d & x\\
a & b & q
\end{Bmatrix}\,=\,
\frac{\delta_{pq}}{(2p+1)}.
\end{equation}
  \item Given the relation (\ref{F43}), the unexpected new
symmetry of the $6j$ symbol  discovered in 1958 by Regge 
\cite{Reg2} (see also \cite{BiLo8,Russi}) is recognized as
a "trivial"  set of permutations on the parameters
$\alpha, \beta$ that leaves
$_4F_3$ invariant. Combining the Regge symmetry and
the "classical" ones, one get a total number of 144 
algebraic symmetries for the $6j$.
Note however that implications  of Regge symmetry
on the geometry of the quantum tetrahedron, 
taken into account  in \cite{Rob}, 
certainly deserve further investigations
also in view of the relevance of this topic in completely different contexts,
 {\em cfr.} for instance \cite{PiHo}.
\item The Askey hierarchy of orthogonal polynomials can be extended
to a q--hierarchy \cite{Askey}, on the top of
which the q--$_4F_3$ polynomial stands.\\
It is worth noting that
the deformation parameter $q$ was originally assumed by physicists to be a real number 
 related to Planck constant $h$ by $q= e^{h}$, and therefore  it is commonly referred
 to as a `quantum' deformation, while the `classical', undeformed Lie group symmetry is
 recovered at the particular value $q=1$.
 However, when dealing with quantum invariants of knots and $3$--manifolds
 formulated in the framework of "unitary" quantum field theory,
 as done in section 2 and 3,
 $q$ is taken to be a complex root of unity, the case $q=1$
 being considered as the "trivial" one.
 We refer to \cite{Qpoly,BiLoh}
for accounts on the theory of q--special functions
and q--tensor algebras. 
  \end{itemize}

\end{document}